# A stabilized chip-scale Kerr frequency comb via a high-Q reference photonic microresonator

Jinkang Lim,[1,*] Shu-Wei Huang,[1] Abhinav K. Vinod,[1] Parastou Mortazavian,[1] Mingbin Yu,[2] Dim-Lee Kwong,[2] Anatoliy A. Savchenkov,[3] Andrey B. Matsko,[3] Lute Maleki,[3] and Chee Wei Wong[1,*]

[1]Mesoscopic Optics and Quantum Electronics Laboratory, University of California, Los Angeles, CA 90095, USA
[2]Institute of Microelectronics, Agency for Science, Technology and Research, Singapore 117865, Singapore.
[3]OEwaves Inc., 465 North Halstead Street, Suite 140, Pasadena, CA 91107, USA
*Corresponding author: jklim001@ucla.edu; cheewei.wong@ucla.edu

We stabilize a chip-scale $Si_3N_4$ phase-locked Kerr frequency comb via locking the pump laser to an independent stable high-$Q$ reference microresonator and locking the comb spacing to an external microwave oscillator. In this comb, the pump laser shift induces negligible impact on the comb spacing change. This scheme is a step towards miniaturization of the stabilized Kerr comb system as the microresonator reference can potentially be integrated on-chip. Fractional instability of the optical harmonics of the stabilized comb is limited by the microwave oscillator used for comb spacing lock below 1 s averaging time and coincides with the pump laser drift in the long term.

An optical frequency comb is a powerful tool for high precision frequency measurements as an unprecedented frequency ruler [1,2]. Optical clock and metrology, direct broadband gas sensing, atomic-molecular spectroscopy, astronomical spectrograph calibration, light detection and ranging (LIDAR), and optical communication, photonic microwave generation benefit from the frequency comb. However, such innovations mainly remain in the laboratory environment because of its size, weight, and power consumption (SWaP). Power-efficient miniaturization of the combs will expand their versatility and deliver them as a field-usable device.

Parametric four-wave mixing combs in a high $Q$ microresonator driven by a continuous wave (cw) laser is a promising solution allowing on-chip integration [3-9]. Since the first demonstration of a Kerr nonlinearity frequency comb in a micro-scale resonator, low-noise Kerr comb formation, comb space uniformity, and its stability have been intensively studied and shown performances starting to approach that of the mode-locked laser frequency combs [4,10-15]. Currently, with the development of fabrication techniques for ultralow loss resonators and dispersion engineering, research focuses on the active stabilization of the Kerr frequency comb, examined in prior mode-locked laser frequency combs. This is important for applications such as atomic clock, optical to rf frequency division, and metrology. For the full comb self-referencing stabilization, the two degree of freedoms, namely the repetition frequency ($f_r$) and the carrier envelope offset (CEO) frequency, have to be detected and controlled. The repetition frequency is generally in the microwave domain such that it can be read out by a fast photodetector but the detection of CEO frequency needs an interferometric method, either $f$-$2f$ or $2f$-$3f$, which requires an octave-spanning spectrum or broad supercontinuum generation. The interferometric detection of CEO frequency also needs high power especially considering harmonic generation efficiencies for the comb teeth at the $f$-$2f$ or $2f$-$3f$. V. Brasch et al. reported a detection and control of CEO frequency using two transfer lasers from a two-third octave Kerr comb [16] and octave-spanning spectra were directly generated from the microresonators by P. Del'Haye et al. and Y. Oakwachi et al. although the comb repetition frequencies were hundreds of GHz [8,14], somewhat harder to detect the repetition frequency directly. Although the supercontinuum can be generated by an external highly nonlinear fiber with a cavity soliton pulse with tens of GHz repetition frequencies, it requires high power amplification owing to the Kerr comb's low pulse energy in nature [17]. In addition, the pulse amplification and nonlinear spectral broadening may induce undesirable phase noise to the comb.

Alternatively, a fully stabilized Kerr comb can be achieved by using a stable optical reference to control the second degree of freedom without detecting CEO frequency and by locking its repetition frequency to a microwave oscillator. This has been realized and obtained a fractional instability of

$7\times10^{-13}$ at 1 s integration [4] assisted by an optical frequency comb reference. Stabilizations using all optical methods such as Rb clock transitions via transfer lasers [18,19] and a probe laser [18] have also been proposed and demonstrated. In this study, we utilize a high-$Q$ MgF$_2$ whispering gallery mode (WGM) reference microresonator to stabilize the pump laser frequency and a microwave oscillator to stabilize the repetition frequency, which simplifies the system. Using this technique, we stabilize the pump laser frequency over a wide detuning range and achieve better than $5\times10^{-11}$ at 1 s integration, limited by the local microwave oscillator used for the comb repetition frequency control.

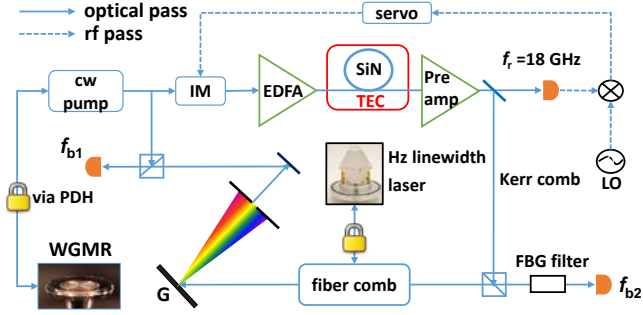

**Fig. 1.** Experimental set-up. The cw pump is coupled into the Si$_3$N$_4$ microring and generates a low-noise phase-locked Kerr comb. The cw pump is stabilized to the high-$Q$ WGM resonator reference via PDH lock. Both the cw pump and comb stabilities are measured by beating them against a fiber laser comb stabilized to a 1 Hz linewidth reference laser. IM: intensity modulator, G: reflection grating, LO: local oscillator, TEC: thermo-electric cooler, FBG: fiber Bragg grating.

Fig. 1 illustrates the experimental setup. The output near 1602 nm from a tunable extended cavity laser diode (ECDL, *New Focus* TLB-6700) is split and one arm is launched into an intensity modulator and amplified by an L-band erbium doped fiber amplifier (EDFA) up to 33.5 dBm. The amplified output is coupled into a chip-scale Si$_3$N$_4$ microring resonator possessing an 18 GHz free spectral range (FSR) with the coupling loss of ~3 dB using an achromatic lens. By detuning the pump laser frequency (red shift), we generate a low noise Kerr comb [12] whose optical spectrum is shown in Fig. 2. We recently demonstrated this comb spacing uniformity, which showed $2.8\times10^{-16}$ relative inaccuracy in the whole comb spectrum [15]. We filter out a part of the comb at the C-band and amplify it to perform an *out-of-loop* measurement by heterodyne beating it against a 250 MHz repetition frequency fiber laser comb (Menlo System) which is referenced to an ultrastable optical reference (Stable Laser Systems) possessing a 1 Hz linewidth and less than 0.1 Hz/s drift. The inset in Fig. 2 shows the fundamental comb beat at a high-speed InGaAs (*EOT* ET-3500F) photodetector. We observe a clear 18 GHz peak without any noticeable noise peak around it implying a low noise phase-locked state. The beam from the other arm is sent to the high-$Q$ WGM microresonator reference on the temperature controlled mount and is coupled by a prism. Then the laser is stabilized by the Pound-Drever-Hall (PDH) locking technique [21]. We modulate the pump laser frequency to find a cavity resonance by applying a triangle modulation voltage to the ECDL current modulation knob that is eventually used to stabilize the laser to the WGM microresonator with ~ 1 MHz bandwidth. The PDH error signal is optimized by controlling the modulation depth and frequency on the phase modulator and coupling power into the resonator for the best laser-WGM microresonator stabilization.

The WGM microresonator from OEwaves is made of single crystalline MgF$_2$ by mechanical polishing and has a 3.45 mm in radius ($r$) and 25 μm rim thickness ($L$). The quality factor was measured to be $2.4\times10^9$. To reduce its thermal expansion sensitivity, the microresonator is sandwiched by a laminating Zerodur structure which reduces the phase noise at the low Fourier frequency regime by a factor of three. After the pump laser stabilization, we measure the frequency instability of both the pump laser and Kerr comb by beating them against a fiber laser reference comb at 1602.7 nm ($f_{b1}$) and at 1557.1 nm ($f_{b2}$) respectively. The fiber comb is filtered out by a reflection grating and combined with the cw pump laser. The generated Kerr comb and the fiber laser comb are mixed and then filtered by a fiber grating filter.

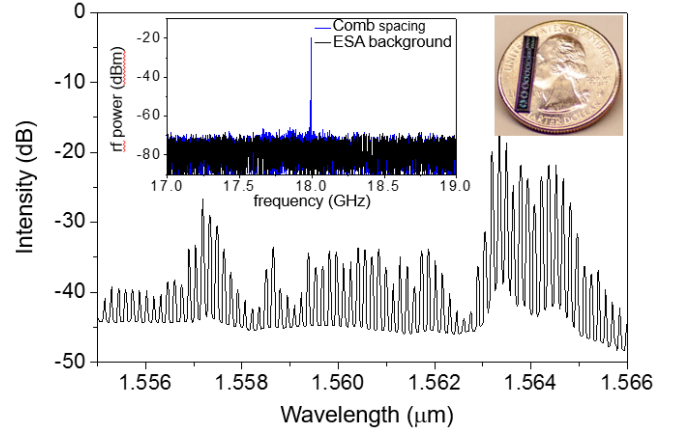

**Fig. 2.** Optical Kerr comb spectrum after a bandpass filter at C-band. Inset (left) is a fundamental beat frequency. A clear 18 GHz peak with signal-to-noise ratio of 50 dB at the fundamental beat frequency without any noticeable noise peak around it implying a low noise phase-locked Kerr comb state. Another inset (right) is the Si3N4 microring on chip.

Fig. 3 shows the single sideband (SSB) phase noise of the stabilized pump laser by beating it against the fiber comb ($f_{b1}$). We achieve -2.5 dBc/Hz at 10 Hz offset frequency in a simple aluminum box shown in Fig. 3 inset. The higher offset frequency noise is dominated by technical noise from the pump laser and the environmental perturbations of the WGM microresonator reference. Peaks in the phase noise measurement stem from the 60 Hz electric power line noise and acoustic noise which can be further reduced by vacuum and thermal isolation of the WGM microresonator. We also calculate and plot the thermorefractive noise limit of our WGM microresonator reference. Due to the small mode volume of the WGM microresonator, the thermorefractive noise is major noise source for our WGM microresonator determining the thermal noise limit. The thermorefractive noise for the cylindrical WGM microresonator geometry is derived and described by [22].

$$S_\varphi^2(f)_{tre} = \frac{k_B \alpha_n^2 T^2}{\rho C V_m} \frac{r^2}{12D} \left[ 1 + \left( \frac{2\pi r^2 |f|}{9\sqrt{3}D} \right)^{3/2} + \frac{1}{6}\left( \frac{r^2}{D} \frac{\pi f}{4m^{1/3}} \right)^2 \right]^{-1}, \quad (1)$$

where $k_B$ is Boltzmann constant, $\alpha_n$ is the thermorefractive coefficient of the material, $\rho$ is the material density, $C$ is the specific heat capacity, $V_m$ is the mode volume of the WGM microresonator mode, $D$ is the temperature diffusion coefficient, and $m$ is the mode order defined by $m=2\pi r n/\lambda$. The stabilized laser phase noise is close to the thermodymanical noise limit at the low offset frequency regime. The values of parameters used in the calculation is $\alpha_n$ = $6\times10^{-7}$/K, $T$ = 300 K, $\rho$ = 3.18 g·cm$^{-3}$, $C$ = $9.2\times10^{6}$ erg·g$^{-1}$·K$^{-1}$, $V_m$ = $2.62\times10^{-6}$ cm$^3$, $D$ = $7.17\times10^{-2}$ cm$^2$·s$^{-1}$, $\lambda$ = 1602.7 nm, and $n$ = 1.37. The phase noise of our stabilized pump laser is close to the thermorefractive noise limit near the carrier frequency regime.

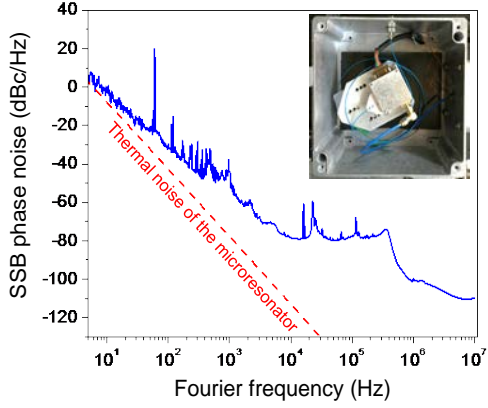

**Fig. 3.** SSB phase noise of the cw pump laser stabilized to the WGM microresonator reference (blue solid line) with the thermorefractive noise limit of the WGM microresonator shown in the red dashed line. The phase noise is measured by heterodyne-beating the pump laser against a Menlo comb tooth at 1602.7 nm. The Menlo comb is stabilized to the 1 Hz laser which has a 0.1 Hz/s drift rate. Inset: the packaged WGM microresonator in an aluminum box. TEC, photodetector, and coupling optics are integrated in a small footprint (40×40×15 mm$^3$).

Furthermore, we fully stabilize the Kerr comb by stabilizing the comb spacing via input power control using a fiber-coupled intensity modulator into the Si$_3$N$_4$ microring resonator. The power modulation provides the comb spacing shift by 8.1 MHz/W. We stabilize the 18 GHz comb repetition frequency to an 18 GHz local oscillator with a feedback bandwidth of 125 kHz as shown in Fig 4. We measure the Kerr comb stability at 1557.1 nm ($f_{b2}$), separated by 304 mode numbers from the pump laser frequency, for the pump laser locked only, and both pump laser and comb repetition frequency locked, respectively, illustrated in Fig. 5. We also plot the fractional instability of the stabilized cw pump laser to see the difference with the Kerr comb stability. The brown circle shows the fractional instability ($\Delta f_{b1}/\nu_{pump}$) of the cw pump laser at 1602.7 nm stabilized to the WGM microresonator. This shows that the fractional instability is $2.3\times10^{-11}$ at 1 s averaging time and goes lower below 1 s averaging time (termed the short-term in this paper), which agrees with the low phase noise measured in Fig. 3. The fractional instability increases with longer averaging times implying that the uncompensated thermo-mechanical loss degrades the quality of the long term stability of the WGM microresonator. The blue diamond shows the stability ($\Delta f_{b2}/\nu_{1557.1}$) of the Kerr comb at 1557.1 nm with the cw pump laser stabilized to the WGM resonator. The stability is improved by two orders of magnitude at 1 s averaging time, compared with the free-running comb (black squares) to $1.29\times10^{-10}$, which is limited by the Si$_3$N$_4$ microring cavity FSR drift. Above the 1 s averaging time, the comb stability follows the stability of the pump laser.

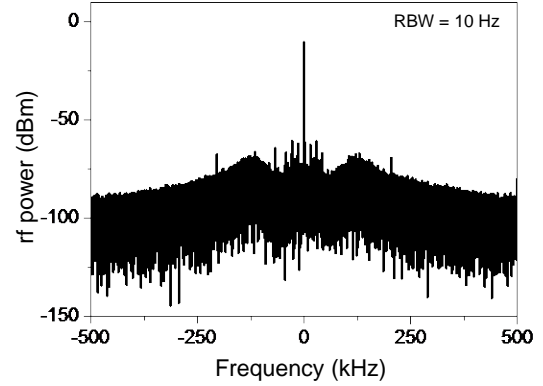

**Fig. 4.** The stabilized 18 GHz repetition frequency with resolution bandwidth (RBW) of 10 Hz measured by an electronic spectrum analyzer. A servo-bump at 125 kHz is observed.

By using the comb equation stabilized to an optical reference laser, the frequency change at the measured mode number can be calculated, assuming the changes in the pump laser frequency and offset frequency ($f_0$) are orthogonal, by $\Delta\nu_{1557.1} = m\Delta\nu_{pump} + \{1-m\}\Delta f_0$, where m = $n_{1557.1}/n_{pump} \approx 1.029$ in our measurement. By putting the comb algebraic relation for the $n_{1557.1}$ comb tooth, $\Delta f_0 = \Delta\nu_{1557.1} - n_{1557.1}\Delta f_r$, the equation can be rewritten with the pump laser stability and the repetition frequency stability,

$$\Delta\nu_{1557.1} = \Delta\nu_{pump} + \left(\frac{m-1}{m}n_{1557.1}\right)\Delta f_r, \quad (2)$$

therefore, the change in the comb spacing frequency has the weighting factor of 301.4 in our comb stability measurement at 1557.1 nm. The long-term stability measurements show that $\Delta\nu_{1557.1} \approx \Delta\nu_{pump}$ implying that the frequency shift at 1557.1 nm largely stems from the microresonator-stabilized pump laser instability. We also measure the $f_r$ shift at 18 GHz due to the the pump laser frequency change. We found the ratio (pump laser frequency detuning versus comb spacing) is 57 Hz/MHz which corresponds to 17.15 kHz/MHz (1.7 %) at 1557.1 nm from Eq. (2) for the mode number, $n_{1557.1}$= 10697 and hence, the pump laser shift induces negligible change to the comb spacing change. However, the short-term stability is far larger than the pump laser stability implying that the noise comes from the comb spacing drift ($\Delta f_r$) caused by the Si$_3$N$_4$ microring resonator. Therefore, we stabilize the 18 GHz comb spacing to an 18 GHz microwave local oscillator to improve the short-term stability. Indeed, the short-term stability is improved (green triangle) and the stability at 1 s averaging time was enhanced by approximately 3 times compared with the pump laser locked to the WGM microresonator reference only. Above 1 s averaging time, the stability remains approximately same with the result achieved with the pump laser locked only, which confirms that the stability coincides with the pump laser drift. Currently, the short-term stability is limited by our 18 GHz local oscillator stability (red line) in Fig. 5. Near 1 s averaging time, the instability measurement shows $4.9\times10^{-11}\,\tau^{0.5}$ implying the random walk frequency noise caused by environmental factors such as vibration, and temperature fluctuations. The instability is slightly higher than that of the local oscillator used for the comb spacing lock, which is attributed to some residual noise of the repetition frequency stabilization near the carrier frequency regime as shown in Fig. 4.

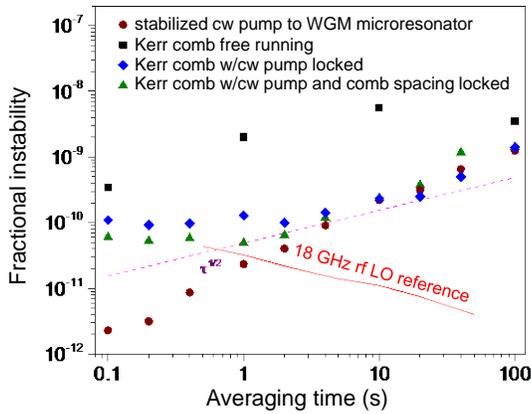

**Fig. 5.** Fractional instability of the stabilized cw pump at 1602.7 nm and Kerr comb at 1557.1 nm. Brown circle indicates the fractional instability of the pump laser frequency stabilized to the WGM microresonator reference. The plot also shows that the fractional instability of the free-running (black squares), the pump laser frequency only stabilized (blue diamond), and both the pump laser frequency and comb repetition frequency stabilized (green triangle) Kerr comb tooth at 1557.1 nm respectively.

In summary, we proposed and demonstrated a fully stabilized on-chip $Si_3N_4$ phase-locked comb using a stable high-$Q$ WGM microresonator reference. This result shows any coherent comb can be similarly stabilized by controlling two degrees of freedom like soliton Kerr combs. We measured the stabilized pump laser phase noise and the Kerr comb stability by heterodyne beating the pump laser and a Kerr comb tooth against a mode-locked fiber laser comb referenced to an ultrastable laser possessing 1 Hz linewidth and 0.1 Hz/s drift. When we stabilized the pump laser to the WGM microresonator reference, the laser shows a phase noise of -2.5 dBc/Hz at 10 Hz and the Kerr comb stability at 1557.1 nm, separated by 304 mode numbers from the pump laser, is improved by two orders of magnitude in 1 s averaging time, to $1.29\times10^{-10}$ and the long-term stability coincides with the pump laser frequency drift. We also observed the coupling between the comb spacing change and the pump laser frequency change (17 kHz/MHz at 1557.1 nm) is insignificant in the pump laser stabilization scheme. The short-term stability of the Kerr comb is limited by the FSR drift of the $Si_3N_4$ microring resonator with the pump laser locked only. When both pump laser and comb repetition frequency are stabilized simultaneously, the short-term stability is limited by the local microwave oscillator used for the comb spacing stabilization. The stability is improved by 3 times in 1 s averaging time resulting in the fractional instability of $4.9\times10^{-11}$. In the future development, the pump laser long-term stability needs to be improved. This can be achieved by better isolation of the WGM microresonator from the environment and better thermal compensation design lowering the thermal expansion sensitivity, which could allow for improved Kerr comb frequency oscillators with a merit in SWaP for various precision measurements.

**Funding and acknowledgement.** The authors acknowledge the funding support from the DARPA Direct On-Chip Digital Optical Synthesizer (DODOS) program under Dr. Robert Lutwak with contract HR0011-15-2-0014, the Air Force Young Investigator award (FA9550-15-1-0081 to S.W.H.), and the Office of Naval Research (N00014-14-1-0041). The authors acknowledge discussions with Jinghui Yang, Yongnan Li, and Hao Liu.